\documentclass[11pt,twoside]{article}

\usepackage[utf8]{inputenc}
\usepackage{amsmath, amssymb, amsfonts, amsthm}
\usepackage{graphicx}
\usepackage{float}
\usepackage{booktabs}
\usepackage{multirow}
\usepackage{arydshln}

\usepackage[colorlinks=true, urlcolor=blue, citecolor=black, linkcolor=black]{hyperref}

\usepackage{tikz, pgfplots}
\pgfplotsset{compat=1.18}
\usepgfplotslibrary{groupplots, fillbetween, external}
\usetikzlibrary{patterns, arrows.meta, positioning, shapes}
\usepackage{pgfplotstable}
\usepackage{subcaption}

\usepackage{enumitem}

\usepackage[authoryear]{natbib}

\usepackage{lmodern}
\usepackage{microtype}
\usepackage{titlesec}
\titleformat{\section}{\normalfont\Large\bfseries}{\thesection}{1em}{}
\titleformat{\subsection}{\normalfont\large\bfseries}{\thesubsection}{1em}{}

\usepackage[font=small,labelfont=bf]{caption}
\usepackage{comment}

\usepackage{geometry}
\title{A Favre-Averaging Shallow Water Framework for Aerated Flows with Friction Factor Decomposition}
\author{Matthias Kramer \\ 
School of Engineering and Technology (SET)\\
The University of New South Wales, Canberra, Australia \\
\texttt{m.kramer@unsw.edu.au}
}
\date{January 2026}

\usepackage{geometry}
\usepackage{fancyhdr}
\usepackage[T1]{fontenc}
\usepackage{newtxtext,newtxmath} 

\usepackage{amsmath, amssymb}

\usepackage{microtype}
\titleformat{\section}
  {\normalfont\normalsize\bfseries}
  {\thesection}{1em}{}

\titleformat{\subsection}
  {\normalfont\normalsize\itshape}
  {\thesubsection}{1em}{}

\titleformat{\subsubsection}
  {\normalfont\normalsize\itshape}
  {\thesubsubsection}{1em}{}

\titlespacing*{\section}{0pt}{12pt plus 2pt minus 2pt}{4pt}
\titlespacing*{\subsection}{0pt}{10pt plus 2pt minus 2pt}{4pt}
\titlespacing*{\subsubsection}{0pt}{6pt plus 1pt minus 1pt}{2pt}

\begin{document}
\maketitle
\thispagestyle{fancy}

\begin{abstract}
Accurate prediction of flow resistance in high-Froude-number aerated flows remains challenging due to air entrainment, which causes strong spatial variability in mixture density. \textcolor{black}{Here, we introduce for the first time a density-weighted (Favre) averaging approach within a Shallow Water Equation framework specifically tailored to account for this strong mixture density variability.} Within this framework, we present a novel Darcy–Weisbach friction factor formulation that decomposes contributions associated with uniform flow, spatially varying flow, and temporally evolving flow, and incorporates momentum and pressure correction factors reflecting the vertical structure of the mixture. \textcolor{black}{Application to experimental data demonstrates that spatial flow development systematically reduces the effective friction factor relative to the uniform-flow estimate, and that momentum-based and energy-based formulations yield nearly identical results. The framework recovers classical uniform-flow predictions in the quasi-uniform downstream region and reduces to standard single-phase formulations in the absence of aeration. Overall, it provides a physically consistent tool for resistance prediction in high-Froude-number spillways, chutes, and open-channel systems, with a structure compatible with depth-averaged numerical solvers.}
\end{abstract}


\section{Introduction}
High-Froude-number open-channel flows often exhibit self-aeration due to strong turbulence and surface instabilities. The entrainment of air modifies the mixture density, turbulence structure, and momentum distribution, which strongly influence hydraulic resistance, energy dissipation, and flow dynamics. Accurately predicting friction in these flows is essential for a wide range of environmental and engineering applications, from river hydraulics to industrial water transport.

\textcolor{black}{Aerated open-channel flows have been studied extensively over the past decades, 
with research focusing on air concentration distribution \citep{Straub1958,  Wood1991, Chanson1992,Bung2009, KramerValero2023}, energy dissipation \citep{Boes03Design,Felder2009,Felder2011a}, and flow 
resistance \citep{Chanson2004, Ruff2002, scheres2020flow,Kramer2021}. Flow resistance in particular 
remains an active area of research, as accurate friction factor predictions are essential 
for the design and safety assessment of spillways, stepped chutes, and other high-velocity open-channel structures. The most commonly used friction factor formulation is of Darcy--Weisbach type 
\citep{Wood1991,Chanson1992,Boes2000,Ruff2002,Chanson2004}}

\begin{equation}
f_e = \frac{8  g d_{\mathrm{eq}} S_f}{\langle \overline{u_w} \rangle^2} = \frac{8  g d_{\mathrm{eq}}^3 S_f}{q^2},
\label{FrictionUniform}
\end{equation}
where \(\langle \overline{u_w} \rangle\) is the mean water velocity, averaged over both flow depth and time, \(g\) is the acceleration due to gravity, $q = \langle \overline{u_w} \rangle \, d_\mathrm{eq} $ is the specific water flow rate, $S_f = \tau_0/(\rho_w g d_\mathrm{eq})$ is the friction slope with $\tau_0$ the bed shear stress and $\rho_w$ the water density, and \(d_\mathrm{eq} \) is the equivalent clear-water depth, defined as

\begin{equation}
d_{\mathrm{eq}} =  \int_{z=0}^{z_{90}} (1 - \overline{c}) \, \text{d}z
\label{EquiClear}
\end{equation}
where \(\overline{c}\) is the time-averaged volumetric air concentration, $z$ is the bed-normal coordinate, measured positive upward from the channel bed, and \(z_{90}\) denotes the mixture flow depth where \(\overline{c} = 0.9\). Here, the operators $\overline{(\cdot)}$ and $\langle \cdot \rangle$ indicate time and spatial averaging, respectively. Equation (\ref{FrictionUniform}) is derived from control volume momentum considerations, where the bed shear stress balances the streamwise momentum loss. In this formulation, the friction slope $S_f$ can alternatively be expressed from an energy perspective, as the head loss per unit length

\begin{equation}
 S_f = S_0 - \frac{\partial H_m}{\partial x}, 
 \label{EqSF}
\end{equation}
where $H_m$ is the mean total head of the mixture referenced to the channel bed elevation $z_0$, measured vertically from a fixed datum, $x$ is the streamwise coordinate along the channel
bed, and $S_0 = - \partial z_0/\partial x$ is the bed slope. For steady, uniform flow conditions, the friction slope equals the bed slope, i.e., $S_f = S_0$. 

Despite its widespread use, (\ref{FrictionUniform}) presents several important limitations. First, it has mostly been applied for uniform flow \citep{Chanson1992,Chanson1993,Boes03Design,Bung2009,wang2022self}, and where non-uniform flows are considered, correction factors accounting for \textcolor{black}{the vertical non-uniformity of velocity profiles} are often neglected \citep{Felder2011a,scheres2020flow,mozer2025determining}. Second, pressure correction factors, which account for the effects of the air–water mixture on the flow, are mostly absent from literature formulations. Third, (\ref{FrictionUniform}) is restricted to steady conditions and does not account for temporal variations of depth or velocity, which may be significant in transient aerated flows.

Thus, there is a clear need for a unifying friction framework that is rigorously derived and accounts for both spatial and temporal variations in aerated flows.

\textcolor{black}{A key challenge in developing such a framework is the strong spatial variability of mixture density due to air entrainment, which renders standard (Reynolds) depth-averaged equations ill-suited: concentration-velocity correlation terms appear that require additional closure assumptions. The General Shallow Water Equations (GSWEs) of \cite{Pokrajac2023} represent a recent generalisation of the SWE framework that accommodates a range of practical applications. However, for high-Froude-number aerated flows, where air entrainment creates a continuously varying mixture density throughout the depth, a density-weighted averaging approach becomes necessary. Favre-averaging, well established in compressible and variable-density turbulence modelling \citep{Favre1965,Favre1969,Wilcox1993}, provides such an approach naturally: it eliminates concentration-velocity correlation terms by construction and yields depth-averaged equations in which the mean water velocity emerges as the primary velocity scale, consistent with existing experimental practice. This motivates the present use of Favre averaging as the foundation for a SWE framework tailored to high-Froude-number aerated flows.}

\textcolor{black}{
In this study, we derive a Darcy–Weisbach friction factor formulation within a density-weighted Shallow Water Equation (SWE) framework. We introduce Favre averaging for aerated flows and the corresponding depth-averaged continuity and momentum equations ($\S$
 \ref{Sec2.1} and \ref{Sec2.2}), followed by an explicit decomposition of the friction factor into distinct physical contributions  ($\S$ \ref{Sec2.3}). The framework is demonstrated using an existing experimental dataset ($\S$ \ref{Sec3}), and its limitations and broader implications are discussed ($\S$ \ref{Sec4}).}

\section{Methods}
\label{Sec2}
Below, we introduce density-weighted (Favre) averaging for aerated mixtures and define the key quantities and correction factors required to describe the non-uniform vertical structure of the flow. These definitions form the basis for the subsequent derivation of the Favre- and depth-averaged continuity and momentum equations.

\subsection{Favre averaging in aerated flows}
\label{Sec2.1}
In aerated flows, the local mixture density can vary significantly due to entrained air. To properly account for the effects of density fluctuations and to preserve conservation of mass and momentum, it is useful to employ Favre averages. 

\subsubsection{Definition of Favre averages}

For an arbitrary quantity $\phi$, the Favre average is defined as \citep{Favre1965,Favre1969}

\begin{equation}
   \tilde{\phi} = \frac{\overline{\rho_m \phi}}{\overline{\rho_m}}, \quad \phi = \tilde{\phi} + \phi'',
\end{equation}
where $\rho_m$ is the instantaneous mixture density, the operator $\tilde{(\cdot)}$ denotes Favre-averaging, and $\phi''$ is the Favre fluctuation, which satisfies by construction

\begin{equation}
   \overline{\rho_m \phi''} = 0.
\end{equation}

For aerated open-channel flows, the time-average of the  mixture density multiplied by a quantity $\phi$ can be expressed as

\begin{equation}
\overline{\rho_m \phi} = \rho_a \, \overline{c \phi} + \rho_w \, \overline{(1-c) \phi} 
\approx \rho_w \, \overline{(1-c) \phi},
\end{equation}
where $c$ is the instantaneous volumetric air concentration, $\rho_a$ and $\rho_w$ are the densities of air and water, respectively. Since $\rho_a \ll \rho_w$, the term $\rho_a \, \overline{c \phi}$ is neglected.

Depth-averaged Favre quantities are obtained by integrating from the channel bed ($z=0$) to the characteristic mixture depth $z_{90}$, following \cite{Wood1991}. Droplet flow above $z_{90}$ is neglected as it contributes negligibly to mass and momentum transport \citep{Chanson1994book,Chanson1996book}. Combining Favre averaging with the density approximation $\overline{\rho_m} \approx \rho_w (1-\overline{c})$ yields a Favre- and depth-averaged quantity

\begin{equation}
\langle \tilde{\phi} \rangle
= \frac{\int_{z=0}^{z_{90}} \overline{\rho_m \phi} \, \mathrm{d}z}{\int_{z=0}^{z_{90}} \overline{\rho_m} \, \mathrm{d}z}
\;\; \approx \;\;
\frac{\rho_w \int_{z=0}^{z_{90}} (1-\overline{c}) \, \overline{\phi} \, \mathrm{d}z}{\rho_w \int_{z=0}^{z_{90}} (1-\overline{c}) \, \mathrm{d}z}
\;=\;
\frac{\int_{z=0}^{z_{90}} (1-\overline{c}) \, \overline{\phi} \, \mathrm{d}z}{d_\mathrm{eq}},
\label{depthFavre}
\end{equation}
where the equivalent clear-water depth appears in the denominator. Equation (\ref{depthFavre}) assumes that fluctuations of $c$ are weakly correlated with $\phi$, which is reasonable for bulk aerated flows. Key Favre-averaged quantities used later in the development of the shallow water framework are:

\begin{itemize}
    \item Mixture mass ($\phi = 1$)
\begin{equation}
\langle \tilde{1} \rangle
= \frac{\int_{z=0}^{z_{90}} \overline{\rho_m}  \, \mathrm{d}z}
       {\int_{z=0}^{z_{90}} \overline{\rho_m} \, \mathrm{d}z} 
\approx \frac{\int_{z=0}^{z_{90}} (1-\overline{c}) \, \mathrm{d}z }{ d_\mathrm{eq} } = 1 \, ,
\label{FavreMass}
\end{equation}

    \item Mixture velocity ($\phi = u_m$)
    \begin{equation}
\langle \tilde{u_m} \rangle
= \frac{\int_{z=0}^{z_{90}} \overline{ \rho_m u_m} \, \mathrm{d}z}
     {\int_{z=0}^{z_{90}} \overline{\rho_m} \, \mathrm{d}z}
\approx
\frac{ \int_{z=0}^{z_{90}} (1-\overline{c}) \, \overline{u_m} \, \mathrm{d}z}
     { d_{\mathrm{eq}} } \, ,
     \label{FavreVel}
\end{equation}

    \item Squared velocity ($\phi = u_m^2$)
   \begin{equation}
\langle \tilde{u_m}^2 \rangle
= \frac{\int_{z=0}^{z_{90}} \overline{\rho_m u_m^2} \, \mathrm{d}z}
       {\int_{z=0}^{z_{90}} \overline{\rho_m} \, \mathrm{d}z}
\approx \frac{\int_{z=0}^{z_{90}} (1-\overline{c}) \, \overline{u_m}^2 \, \mathrm{d}z}{d_\mathrm{eq}} \, ,
     \label{FavreVel2}
\end{equation}
\end{itemize}
\textcolor{black}{where $u_m$ denotes the local instantaneous streamwise 
mixture velocity.}

It is important to note that the Favre- and depth-averaged mixture velocity  $\langle\tilde{u_m} \rangle$ corresponds to the mean water velocity, i.e., $\langle \tilde{u_m} \rangle \approx \langle \overline{u_w} \rangle$, which naturally follows from the density weighting approach. In the context of $\langle \tilde{u_m}^2 \rangle$, the momentum correction factor $\beta$, which accounts for the vertical non-uniformity of the velocity and mixture density, is then

\begin{equation}
\beta = \frac{\langle \tilde{u_m}^2 \rangle}{\langle \tilde{u_m} \rangle^2} 
= \frac{\int_{z=0}^{z_{90}} (1-\overline{c}) \, \overline{u_m}^2 \, \mathrm{d}z}{d_\mathrm{eq} \, \langle \tilde{u_m} \rangle^2}.
\label{EqBeta}
\end{equation}

\subsubsection{Depth-integrated  pressure and Favre centroid}

Under the hydrostatic assumption, the vertical momentum balance reads

\begin{equation}
\frac{\partial \overline{p_m}}{\partial z} = - \overline{\rho_m} \,  g \cos \theta,
\label{PressureHydro}
\end{equation}

where $\overline{p_m}$ is the mixture pressure and $\theta =\arcsin(S_0)$ is the bed slope angle. Approximating the mixture density as $\overline{\rho_m} \approx \rho_w (1-\overline{c})$ and integrating from $z$ to the mixture depth $z_{90}$, where $\overline{p_m}(z_{90}) = 0$, gives

\begin{equation}
\overline{p_m}(z) = \rho_w g \cos \theta \int_{z}^{z_{90}} (1-\overline{c}(\zeta)) \, \mathrm{d}\zeta,
\end{equation}
where $\zeta$ is a dummy variable. The depth-integrated pressure force per unit width, which appears in the depth-averaged momentum equation, is

\begin{equation}
\int_{z=0}^{z_{90}} \overline{p_m} \, \mathrm{d}z = \rho_w g \cos \theta \int_{z=0}^{z_{90}} \left[ \int_z^{z_{90}} (1-\overline{c}(\zeta)) \, \mathrm{d}\zeta \right] \mathrm{d}z.
\end{equation}

Introducing the Favre centroid of the mixture

\begin{equation}
\langle \tilde{z} \rangle = \frac{\int_{z=0}^{z_{90}} (1-\overline{c}) \, z \,\mathrm{d}z}{d_\mathrm{eq}},
\end{equation}
the depth-integrated pressure can be written compactly as

\begin{equation}
\int_{z=0}^{z_{90}} \overline{p_m} \, \mathrm{d}z = \rho_w g \cos \theta \, d_\mathrm{eq} \, \langle \tilde{z} \rangle.
\label{Pressure}
\end{equation}

To make the analogy with the classical hydrostatic term $\frac{1}{2} \rho_w g \cos \theta \, d_\mathrm{eq}^2$, we define the pressure correction factor $\Omega$ as

\begin{equation}
\Omega = \frac{2 \langle \tilde{z} \rangle }{d_\mathrm{eq}} =  \frac{\int_{z=0}^{z_{90}} (1-\overline{c}) \, z \, \mathrm{d}z}{\frac{1}{2} d_\mathrm{eq}^2},
\label{EqOmega}
\end{equation}

so that the depth-integrated pressure can also be written in the familiar form

\begin{equation}
\int_{z=0}^{z_{90}} \overline{p_m} \, \mathrm{d}z = \frac{1}{2} \rho_w g \cos \theta \, \Omega \,d_\mathrm{eq}^2.
\end{equation}

\subsection{Density-weighted shallow-water framework for aerated flows}
\label{Sec2.2}
\subsubsection{Mixture continuity equation}

Aerated open-channel flows are gravity-driven flows over solid boundaries, where friction opposes the motion induced by gravity. Assuming that the flow is predominantly two-dimensional, the local Favre-averaged continuity equation is expressed as \citep{Wilcox1993,GatskiBonnet2009}

\begin{equation}
\frac{\partial \overline{\rho_m}}{\partial t} + \frac{\partial}{\partial x} (\overline{\rho_m} \tilde{u_m}) + \frac{\partial}{\partial z} (\overline{\rho_m} \tilde{w}_m) = 0,
\label{Eq22}
\end{equation}
with \(\tilde{u_m}\) and \(\tilde{w}_m\) are the Favre-averaged mixture velocities in the streamwise and vertical directions, respectively. Integrating (\ref{Eq22}) from the channel bed ($z = 0$) to the characteristic mixture depth $z_{90}$, and applying the Leibniz rule together with the free-surface kinematic boundary condition \(\tilde{w}_m(z_{90}) = \partial z_{90}/\partial t + \tilde{u_m}(z_{90}) \partial z_{90}/\partial x\), yields 

\begin{equation}
\frac{\partial}{\partial t} \int_{z=0}^{z_{90}} \overline{\rho_m} \, \text{d}z + \frac{\partial}{\partial x} \int_{z=0}^{z_{90}} \overline{\rho_m} \tilde{u_m} \, \text{d}z = 0.
\label{2.3}
\end{equation}

Using depth-averaged Favre quantities defined in (\ref{FavreMass}) and (\ref{FavreVel}), the depth-integrated mixture mass and momentum are approximated as

\begin{equation}
\int_{z=0}^{z_{90}} \overline{\rho_m} \, \text{d}z \approx \rho_w  d_\text{eq},
\label{2.4}
\end{equation}

\begin{equation}
\int_{z=0}^{z_{90}} \overline{\rho_m} \tilde{u_m} \, \text{d}z  \approx \rho_w d_\mathrm{eq}\langle \tilde{u_m} \rangle.  
\label{2.5}
\end{equation}

Substituting (\ref{2.4}) and (\ref{2.5}) into (\ref{2.3}) and dividing by $\rho_w$ gives

\begin{equation}
\frac{\partial d_\mathrm{eq}}{\partial t} + \langle \tilde{u_m} \rangle \frac{\partial d_\mathrm{eq}}{\partial x} + d_\mathrm{eq} \frac{\partial \langle \tilde{u_m} \rangle}{\partial x} = 0,
\label{2.6}
\end{equation}
where the derivative \(\partial ( d_\mathrm{eq} \langle \tilde{u_m} \rangle ) / \partial x\) has been expanded. Equation (\ref{2.6}) describes depth-averaged mass conservation for flows with spatial and temporal variability and forms a key component of the density-weighted shallow water framework for aerated flows.

\subsubsection{Streamwise mixture momentum equation}
The two-dimensional streamwise component of the Favre-averaged mixture equation reads \citep{Wilcox1993,GatskiBonnet2009}

\begin{equation}
\begin{split}
\frac{\partial}{\partial t} (\overline{\rho_m} \tilde{u_m}) + \frac{\partial}{\partial x} (\overline{\rho_m} \tilde{u_m}^2) + \frac{\partial}{\partial z} (\overline{\rho_m} \tilde{u_m} \tilde{w}_m) \\
= \overline{\rho_m} g_x - \frac{\partial \overline{p_m}}{\partial x} + \frac{\partial \overline{\tau_{xx}}}{\partial x} + \frac{\partial \overline{\tau_{xz}}}{\partial z} 
- \frac{\partial}{\partial x} \left(\overline{\rho_m u^{'' 2}_m} \right)
 - \frac{\partial}{\partial z} \left(\overline{\rho_m u''_m w''_m} \right),
\end{split}
\label{EqMomentum}
\end{equation}
where $\tau_{xx}$ and $\tau_{xz}$ are the viscous normal and shear stresses, \(g_x = g S_0\) is the streamwise component of gravitational acceleration associated with the bed slope, and \(u''_m, w''_m\) are the streamwise and vertical Favre velocity fluctuations, respectively. Integrating (\ref{EqMomentum}) over the flow depth \(z\in[0, z_{90}]\) and applying the Leibniz rule, together with the bed boundary condition \(\tilde{w}_m(0)=0\) and the free-surface kinematic condition, yields the depth-integrated momentum balance

\begin{equation}
\frac{\partial}{\partial t} \int_{z=0}^{z_{90}} \overline{\rho_m} \tilde{u_m} \, \text{d}z + \frac{\partial}{\partial x} \int_{z=0}^{z_{90}} \overline{\rho_m} \tilde{u_m}^2 \, \text{d}z
= \int_{z=0}^{z_{90}} \overline{\rho_m} g_x \, \text{d}z - \frac{\partial}{\partial x} \int_{z=0}^{z_{90}} \overline{p_m} \, \text{d}z - \tau_0,
\label{Eq2.8}
\end{equation}
where the bed shear stress $\tau_0$ arises from vertical viscous and turbulent fluxes

\begin{equation}
\tau_0 =  -\int_{z=0}^{z_{90}} \frac{\partial }{\partial z} \left( \tau_{xz} - \overline{\rho_m u''_m w''_m} \right) \text{d}z.
\end{equation}

Note that streamwise viscous and turbulent stress gradients are neglected in the depth-integrated equation, as their contributions are small compared to the corresponding vertical fluxes in shallow, high-Froude-number, gravity-driven flows, consistent with classical shallow-water formulations. 

The depth-integrated mixture momentum and the mixture momentum flux are closed using Favre-averaged quantities defined in (\ref{FavreVel}) and (\ref{FavreVel2}), while the depth-integrated pressure is evaluated using (\ref{Pressure}). Accordingly,

\begin{equation}
\int_{z=0}^{z_{90}} \overline{\rho_m} \tilde{u_m} \, \mathrm{d}z
\approx \rho_w  d_{\mathrm{eq}} \langle \tilde{u}_m \rangle,
\label{Eq2.10}
\end{equation}

\begin{equation}
\int_{z=0}^{z_{90}} \overline{\rho_m} \tilde{u_m}^2  \mathrm{d}z
\approx \rho_w \, d_{\mathrm{eq}} \, \beta \, \langle \tilde{u}_m \rangle^2,
\label{Eq2.11}
\end{equation}

and, assuming hydrostatic conditions,

\begin{equation}
\int_{z=0}^{z_{90}} \overline{p_m} \, \mathrm{d}z
\approx \frac{1}{2} \rho_w g \, \cos\theta \, \Omega \, d_{\mathrm{eq}}^2,
\label{Eq2.13}
\end{equation}
where $\beta$ and $\Omega$ are the momentum and pressure correction factors defined in (\ref{EqBeta}) and (\ref{EqOmega}), respectively. Similar air–water correction factors have been proposed previously \citep{Ohtsu2004, chanson2024momentum}, and their relationship to the present analysis is discussed in $\S$ \ref{CorrFactors}.

The depth-integrated gravitational forcing term is evaluated using the approximation $\overline{\rho_m} \approx \rho_w (1-\overline{c})$, which yields

\begin{equation}
\int_{z=0}^{z_{90}} \overline{\rho_m} g_x \, \mathrm{d}z
\approx \rho_w \int_{z=0}^{z_{90}} (1-\overline{c}) \, g S_0 \, \mathrm{d}z
= \rho_w g d_{\mathrm{eq}} S_0.
\label{Eq2.12}
\end{equation}

Combining (\ref{Eq2.8}) with (\ref{Eq2.10}) to (\ref{Eq2.12}) and dividing by $\rho_w$, leads to the depth-averaged streamwise momentum equation of the density-weighted shallow water framework for aerated flows

\begin{equation}
\frac{\partial (d_{\mathrm{eq}} \langle \tilde{u_m} \rangle)}{\partial t}    + \frac{\partial ( \beta \, d_{\mathrm{eq}} \langle \tilde{u_m} \rangle^2)}{\partial x}= g d_{\mathrm{eq}} S_0 - \frac{\partial}{\partial x}\left( \frac{1}{2} g \, \Omega \, d_{\mathrm{eq}}^2 \cos{\theta}\right) - \frac{\tau_0}{\rho_w}.  \label{Eq2.15}
\end{equation}

\subsection{Effective friction factor formulation}
\label{Sec2.3}

\subsubsection{General decomposition of the effective friction factor}

To derive an explicit expression for the friction factor, closure of the depth-averaged momentum equation requires a representation of the bed shear stress. For aerated open-channel flows, the bed shear stress is expressed in Darcy–Weisbach form as

\begin{equation}
\tau_0 = \frac{1}{8} \rho_w f_e \langle \tilde{u_m} \rangle^2,
\label{EqClosure}
\end{equation}
where \(f_e\) is the effective friction factor of the aerated flow and \(\langle \tilde{u_m} \rangle\) is the Favre-averaged mixture velocity. \textcolor{black}{This closure is general and accommodates both smooth and rough bed conditions, including micro-rough and macro-rough surfaces, through appropriate parametrisation of $f_e$, in the same manner as classical single-phase Darcy–Weisbach formulations.}
Substituting (\ref{EqClosure}) into (\ref{Eq2.15}), and performing the algebraic manipulations outlined in Appendix \ref{AppendixA} yields the explicit decomposition

\begin{equation}
\begin{aligned}
f_e &= 
\overbrace{\frac{8 g d_\mathrm{eq} S_0}{\langle \tilde{u_m} \rangle^2}  \vphantom{\left(
- \frac{\Delta \left( \frac12 g\,\Omega\,d_\mathrm{eq}^2 \cos\theta \right)}{\Delta x} 
- \frac{\Delta (\beta\,d_\mathrm{eq}\langle \tilde{u_m} \rangle^2)}{\Delta x} 
\right)}}^{\text{uniform flow}}
+ 
\overbrace{\frac{8}{\langle \tilde{u_m} \rangle^2} 
\left(
- \frac{\Delta \left( \frac12 g\,\Omega\,d_\mathrm{eq}^2 \cos\theta \right)}{\Delta x} 
- \frac{\Delta (\beta\,d_\mathrm{eq}\langle \tilde{u_m} \rangle^2)}{\Delta x} 
\right)}^{\text{spatially varying flow}}
\\[2mm]
&\quad + 
\underbrace{\frac{8}{\langle \tilde{u_m} \rangle^2} 
\left(- \frac{\Delta (d_\mathrm{eq} \langle \tilde{u_m} \rangle)}{\Delta t} \right)}_{\text{temporally varying flow}},
\end{aligned}
\label{eq:fe_decomp}
\end{equation}
where spatial and temporal derivatives are expressed as finite differences, allowing direct evaluation from measured or simulated depth and velocity profiles. 
Equation~(\ref{eq:fe_decomp}) shows that $f_e$ is an effective quantity incorporating flow development and aeration-induced pressure effects.
The first term recovers the classical Darcy–Weisbach expression for uniform, steady flow, while the second and third terms represent spatial and temporal departures from uniformity, respectively, modified by the pressure and momentum correction factors $\Omega$ and $\beta$. 
\textcolor{black}{Note that these labels describe the flow conditions under which each term dominates or vanishes, rather than implying a superposition of distinct flow types; the decomposition applies to the scalar friction factor $f_e$ and follows directly from the depth-averaged momentum equation.}

For steady, spatially developing flows, the temporal term vanishes. Further assuming gradual spatial variations in $\beta$ and $\Omega$, the friction factor reduces to the Gradually Varied Flow (GVF) form (Appendix~\ref{AppendixA})
\begin{equation}
f_e=\frac{8}{Fr^2}
\left(
S_0
+ \frac{\Delta d_\mathrm{eq}}{\Delta x}
\left(\beta\,Fr^2-\Omega \cos\theta\right)
\right),
\label{Eq36}
\end{equation}
with the Froude number defined as $Fr = \langle \tilde{u_m} \rangle / \sqrt{g \, d_\mathrm{eq}}$.
Equation~(\ref{Eq36}) provides a practical and explicit expression for the friction factor in steadily developing, gradually varied aerated flows, while retaining corrections associated with aeration and \textcolor{black}{vertically} non-uniform velocity profiles.

An alternative GVF friction factor formulation can be derived from energy considerations, yielding an expression identical to the momentum-based formulation (Appendix \ref{feEnergy}). In this energy-based approach, the correction factors in (\ref{Eq36}) are replaced by the kinetic energy correction factor $\alpha$ and the energy-based pressure correction factor $\Omega_E$, which include an additional Favre-averaged velocity term (see Appendix \ref{feEnergy}). 
\textcolor{black}{As a general rule, the energy-based formulation should be used when quantities related to energy fluxes are of interest, such as energy dissipation or head losses, while the momentum-based formulation is more appropriate when estimating forces, pressure effects, or other momentum-related quantities in the flow.
In practice, however, the differences between the two approaches are typically small, as illustrated in the comparison presented in $\S$ \ref{Sec32}}.

\subsubsection{Flow-region interpretation and limiting behaviour}

\begin{figure}[h!]
\centering
\includegraphics[width=0.85\textwidth]{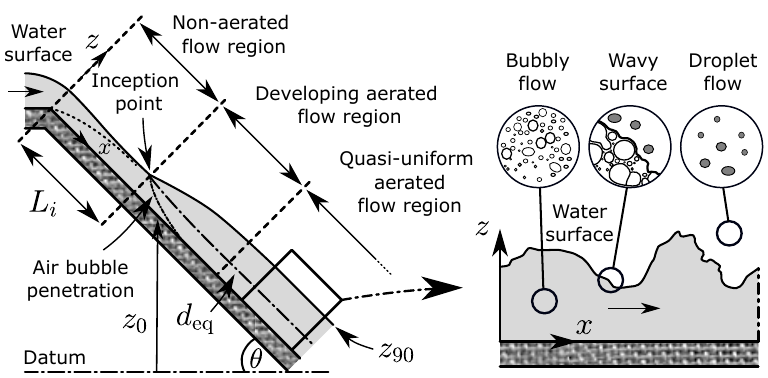}
\caption{Flow regions and flow structure of an aerated high-Froude-number flow; conceptual sketch expanded from \protect \cite{Cain1978}; $d_\mathrm{eq} =$ equivalent clear-water depth; $L_i =$ upstream distance to aeration inception; $x =$ streamwise coordinate; $z=$ bed-normal coordinate; $z_0 =$ bed elevation relative to datum; $z_{90} =$ mixture flow depth; $\theta =$ bed slope angle.}
\label{Fig0}
\end{figure}

The GVF form of the effective friction factor, equation (\ref{Eq36}), can be interpreted in terms of three characteristic streamwise flow regions observed in high-Froude-number aerated flows: a non-aerated clear-water region, a developing aerated  flow region, and a downstream quasi-uniform aerated region (figure \ref{Fig0}). These regions provide a conceptual framework for understanding how the different terms in the friction factor contribute under varying flow conditions.

In the non-aerated flow region, air entrainment is absent ($\overline{c}=0$), and the flow depth equals the equivalent clear-water depth ($d = d_\mathrm{eq} = z_{90}$). The Favre-averaged mixture velocity reduces to the water velocity, $\tilde{u_m} = \overline{u_w}$. Consequently, the momentum correction factor simplifies to its single-phase form
\[
\beta = \frac{\int_{z=0}^{d}  \, \overline{u_w}^2 \, \mathrm{d}z}{d \, \langle {\overline{u_w}} \rangle^2},
\]
and the pressure correction factor becomes $\Omega = 1$. In this limit, (\ref{Eq36}) recovers the classical single-phase GVF friction formulation \citep{HagerBlaser1998}. For rapidly accelerating flows, boundary-layer-based approaches may be used to estimate the friction factor \citep{CastroOrgazHager2009}.

In the developing aerated flow region, the equivalent clear-water depth $d_\mathrm{eq}$ decreases along the streamwise direction, and the full expression of  (\ref{Eq36}) must be applied. Here, the spatial-development term, proportional to $\Delta d_\mathrm{eq}/\Delta x \, (\beta Fr^2 - \Omega \cos \theta)$, plays a dominant role in determining the effective friction factor.

In the downstream quasi-uniform aerated region, the approximated gradient of the equivalent clear-water depth vanishes ($\Delta d_\mathrm{eq}/\Delta x \approx 0$), and (\ref{Eq36}) attains its limiting form for uniform aerated flow, $f_e = 8 g d_\mathrm{eq} S_0 / \langle \tilde{u_m}\rangle^2 = 8 S_0/Fr^2 $. 

In practice, developing aerated flow is always present downstream of the aeration inception (figure \ref{Fig0}), while fully uniform aerated flow occurs only if the chute is sufficiently long. Non-aerated flow exists only upstream of the inception point. The proposed GVF formulation thus provides a practical and physically consistent tool for interpreting and quantifying the spatially varying friction factor across the different flow regions in high-Froude-number aerated flows. For unsteady aerated flows, the temporally varying term from (\ref{eq:fe_decomp}) accounts for temporal changes in depth and velocity.

\section{Results: Application of the proposed framework}
\label{Sec3}
To demonstrate the practical utility of the proposed friction factor framework, 
we apply the developed GVF formulation to a previous experimental dataset from 
\citet{Bung2009}, selectively complementing it with other literature datasets for analyses that do not require detailed velocity measurements.

\begin{figure}[h]
    \centering
    \begin{subfigure}{0.45\textwidth}
        \centering        \includegraphics[height=5cm]{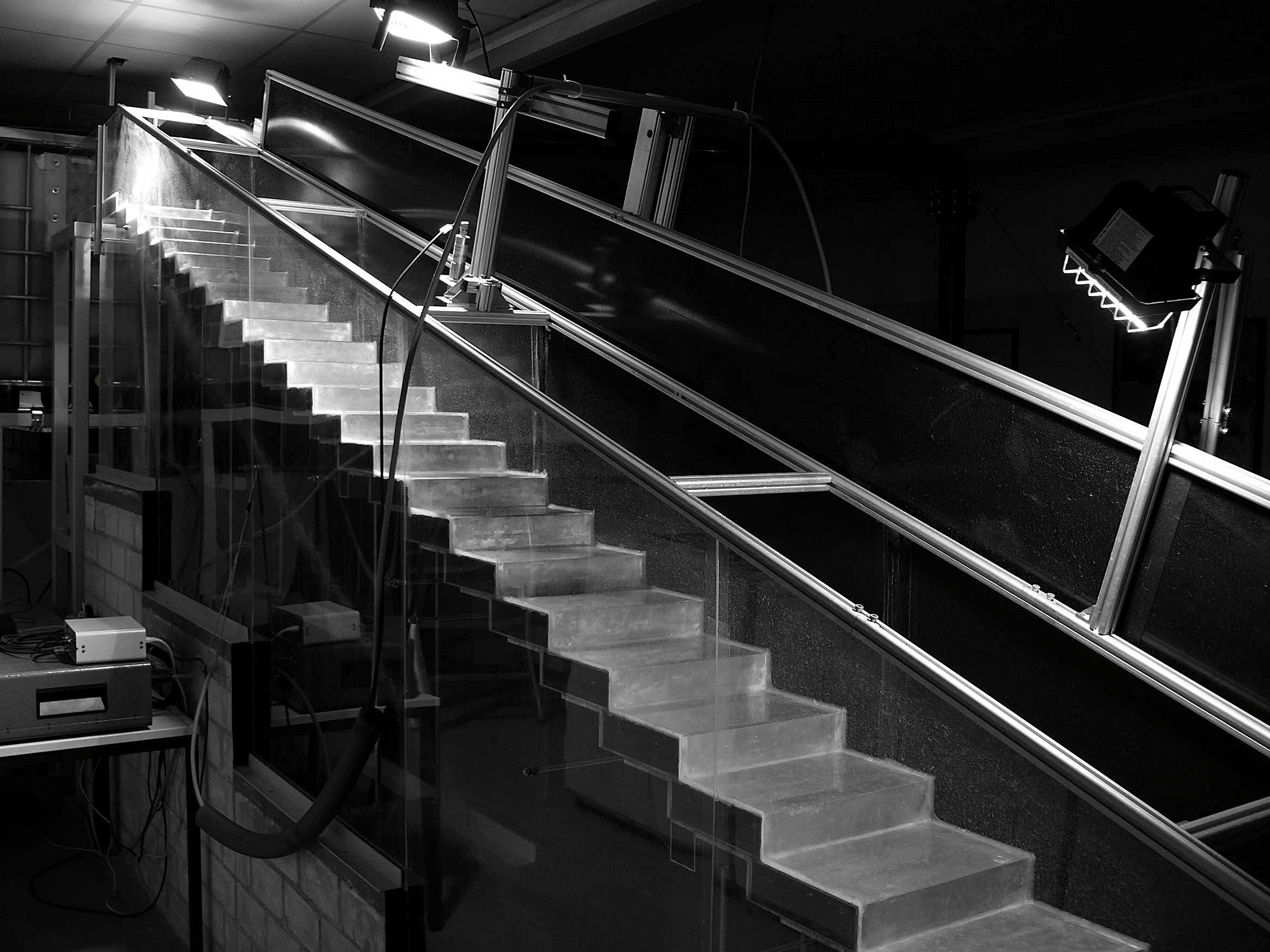}
        \label{fig:left}
    \end{subfigure}
    \begin{subfigure}{0.45\textwidth}
        \centering
        \includegraphics[height=5cm]{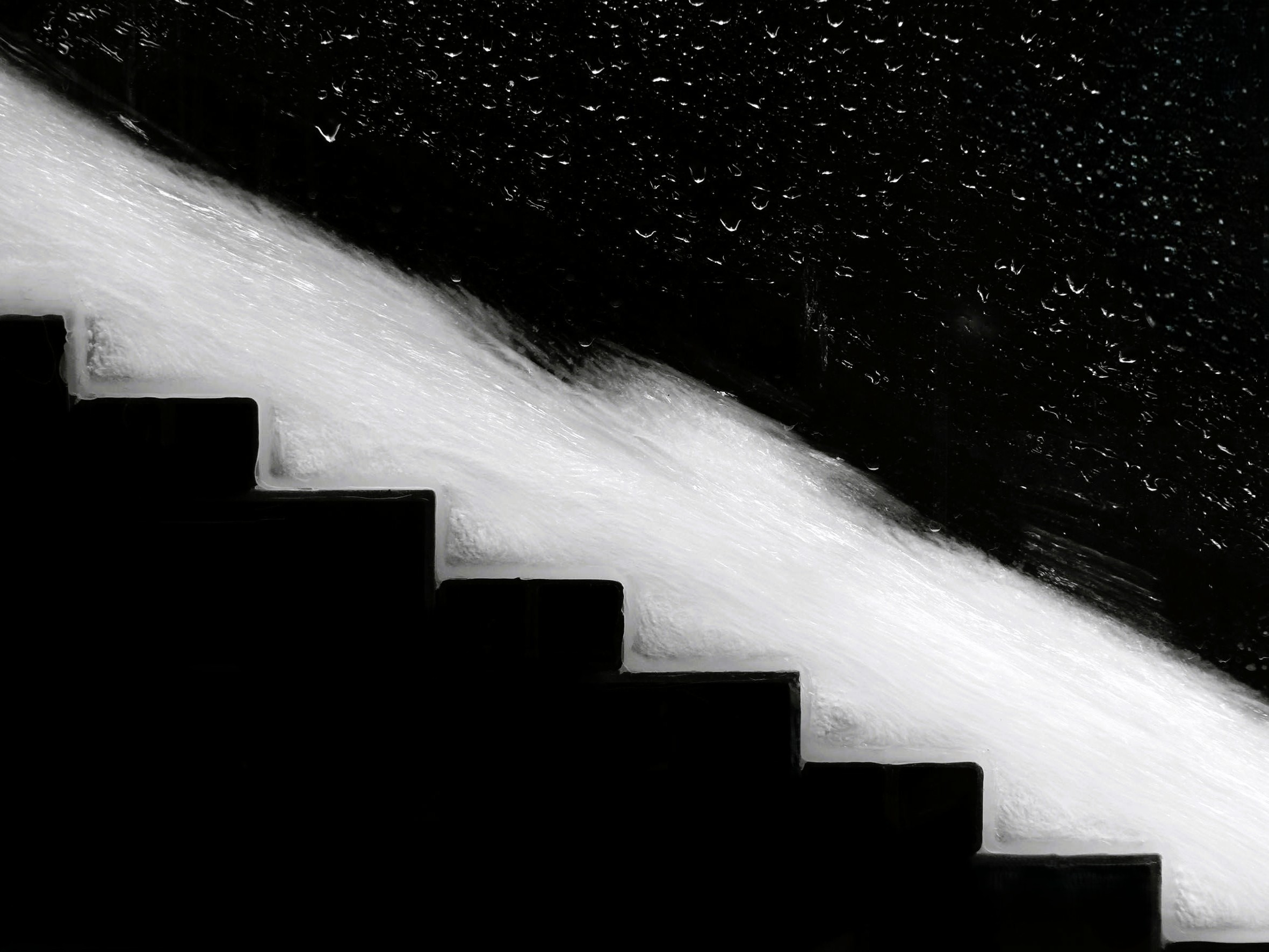}
        \label{fig:right}
    \end{subfigure}
\caption{Experiments from \citet{Bung2009}: (\textbf{a}, left) Experimental setup with $\theta = 13.4^\circ$, $s = 6$ cm, no flow; (\textbf{b}, right) Self-aerated flows down a stepped chute with $q_w = 0.11$ m$^2$/s, $\theta = 26.6^\circ$, $s = 3$ cm. Figures reproduced with permission from \citet{Bung2009}.}
    \label{fig:Bung}
\end{figure}

\textcolor{black}{The experiments of \cite{Bung2009} were conducted on a stepped chute with 
horizontal steps, a channel width of $ 0.30$~m, two bed slopes 
($\theta = 18.4^\circ$ and $\theta = 26.6^\circ$), and two step heights ($s = 3$~cm 
and $s = 6$~cm), illustrated in figure \ref{fig:Bung}. Air concentration 
and interfacial velocity profiles were measured at each step edge downstream 
of the aeration inception point using a dual-tip phase-detection probe. 
The correction factors $\beta$, $\alpha$, $\Omega$, and $\Omega_E$ are 
evaluated using the complete dataset ($3.9 \leq Fr \leq 6.5$), while the 
streamwise evolution of $f_e$ is demonstrated for the specific configuration 
$\theta = 18.4^\circ$, $s = 0.06$~m, $q = 0.11$~m$^2$/s.}

\subsection{Air-water flow properties and correction factors}
\label{CorrFactors}

\begin{figure}[h!]
\centering
\includegraphics{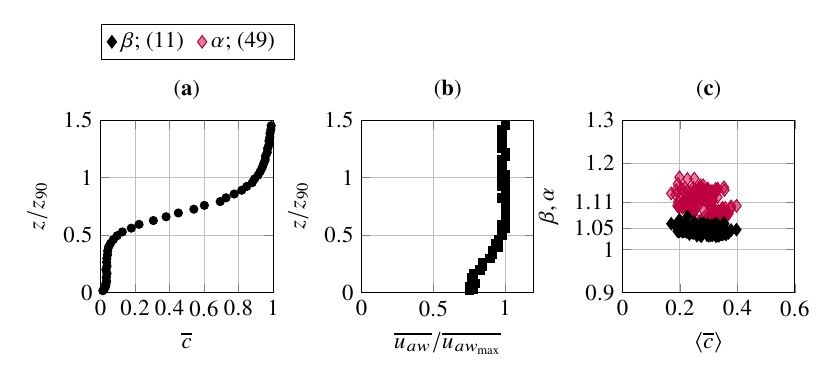}
\caption{Air–water flow properties, momentum and kinetic energy correction factors for aerated flows (data from \cite{Bung2009}): (\textbf{a}) Representative air concentration profile for $Fr = 4.0$, $q = 0.11$ m$^2$/s, $\theta = 18.4^\circ$, step height $s =0.06$ m, step edge 11; (\textbf{b}) Corresponding interfacial velocity profile; (\textbf{c}) Momentum correction factors $\beta$ and kinetic energy correction factors $\alpha$ evaluated for the complete dataset of \cite{Bung2009}, with $3.9 \leq Fr \leq 6.5$.}
\label{Fig1}
\end{figure}

Figure~\ref{Fig1} presents representative vertical profiles of the air concentration $\overline{c}$ and the interfacial velocity $\overline{u_{aw}}$. The air concentration profile exhibits a typical S-shaped distribution (figure \ref{Fig1}\textbf{a}), whose physical origin has been explained by the two-state convolution approach \citep{KramerValero2023,kramer2024turbulent}. The interfacial velocity $\overline{u_{aw}}$ represents the motion of both water and air immediately adjacent to the interface and is approximately equal to the mixture velocity ($\overline{u_{aw}}  \approx \overline{u_m}$), allowing it to be used in depth-averaged momentum calculations. The velocity profile approximately follows a power-law variation through the bulk of the flow, with nearly constant velocity in the upper wavy and spray regions (figure \ref{Fig1}\textbf{b}). Both the air concentration and interfacial velocity profiles were measured using a dual-tip phase-detection intrusive probe. Details are provided by \cite{Bung2009}. 

The momentum correction factor \(\beta\) and the kinetic energy correction factor \(\alpha\) for aerated flows arise from Favre averaging and are defined in (\ref{EqBeta}) and (\ref{EqAlpha}), respectively, to account for the effects of \textcolor{black}{vertically} non-uniform velocity profiles on the mixture momentum and kinetic energy fluxes. Figure \ref{Fig1}\textbf{c}  presents $\beta$ and $\alpha$ for the complete dataset of \cite{Bung2009} as functions of the depth-averaged air concentration, defined as
\begin{equation} 
\langle \overline{c} \rangle = \frac{1}{z_{90}} \int_{z=0}^{z_{90}} \overline{c} \, \text{d}z.   
\end{equation}

The calculations show that $\beta \approx 1.05$ and $\alpha \approx 1.11$, with a slight decrease for increasing $\langle \overline{c} \rangle$, suggesting that vertical velocity gradients become less pronounced as aeration increases. For comparison, \citet[\textcolor{black}{Table~2}]{chanson2024momentum} estimated momentum and kinetic energy correction coefficients using an interfacial velocity power-law model and void fraction profiles governed by an advective–diffusion equation, \textcolor{black}{reporting 
$\beta \approx 1.00$ to $1.04$ and $\alpha \approx 1.04$ to $1.10$ for 
$\theta = 45^\circ$ and $0 < C_\mathrm{mean} < 0.6$, consistent with the values obtained here.}

\begin{figure}[h!]
\centering
\includegraphics{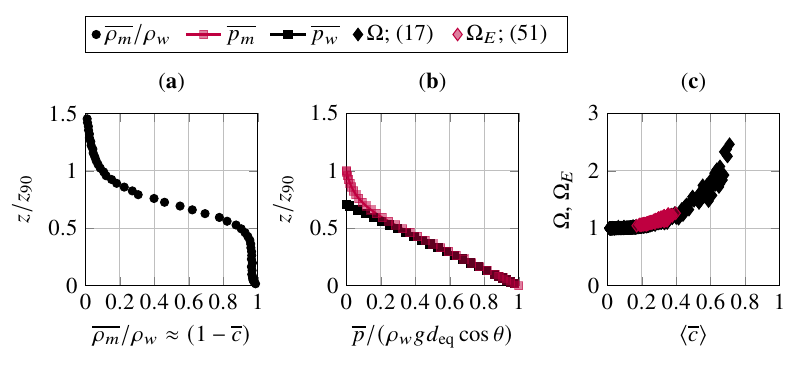}
\caption{Mixture density profile, pressure distribution, and pressure correction factors $\Omega$ and $\Omega_E$ for aerated flows: (\textbf{a}) Representative mixture density profile for $Fr = 4.0$, $q = 0.11$ m$^2$/s, $\theta = 18.4^\circ$, step height $s = 0.06$ m, step edge 11 (data from \cite{Bung2009}); (\textbf{b}) Corresponding mixture pressure and clear-water pressure distributions; (\textbf{c}) Pressure correction factors for micro-rough and macro-rough inverts. $\Omega$ data from \cite{Straub1958}, \cite{Killen1968}, \cite{Bung2009}, \cite{Severi2018}, \cite{Zhang2017DISS}, and \cite{Kramer2018Transiton}; $\Omega_E$ data from \cite{Bung2009}.}
\label{Fig2}
\end{figure}

Figure~\ref{Fig2} shows representative density and pressure characteristics of an aerated flow. Panel \textbf{a} presents the mixture density $\overline{\rho_m}(z)$ normalized by the water density $\rho_w$, while panel \textbf{b} shows the corresponding depth distributions of the mixture pressure ($\overline{p_m}$) and the equivalent clear-water pressure ($\overline{p_w}$). The mixture density exhibits a mirrored S-shaped profile, directly reflecting the vertical variation of water concentration through the relation $\overline{\rho_m}/\rho_w \approx (1 - \overline{c})$. The mixture pressure $\overline{p_m}$ begins at atmospheric pressure at the mixture depth $z_{90}$, consistent with the standard treatment of aerated flows as a continuous fluid  for $0 \leq z \leq z_{90}$ with depth-varying density \citep{Wood1991,Ohtsu2004,chanson2024momentum}. At the channel bed, the pressure equals $\overline{p_m}(z=0) = \overline{p_w}(z=0) = \rho_w g d_\mathrm{eq} \cos \theta$, ensuring consistency between the mixture pressure profile and the equivalent depth representation.

Finally, momentum-based pressure correction factors, $\Omega$, as defined in \eqref{EqOmega}, are derived from 571 air concentration profiles for channels with micro- and macro-rough inverts \citep{Straub1958, Killen1968, Bung2009,Severi2018,Zhang2017DISS,Kramer2018Transiton}. Energy-based correction factors, $\Omega_E$ [equation \eqref{EqOmegaE}], are available only for the 179 profiles from \citet[ $3.9 \leq Fr \leq 6.5$]{Bung2009}, since their computation requires detailed velocity measurements. Figure~\ref{Fig2}\textbf{c} presents the computed pressure correction factors as a function of the depth-averaged air concentration. The results show that the simpler momentum-based factor $\Omega$ closely approximates $\Omega_E$, being about 97 \% of $\Omega_E$. It also increases with $\langle \overline{c} \rangle$, indicating that the pressure correction factor is largely controlled by the depth-averaged air concentration rather than by bed slope or channel geometry, consistent with \citet{chanson2024momentum}.

\subsection{Streamwise development of effective friction factors}
\label{Sec32}
With the momentum correction factor $\beta$ and the pressure correction factor $\Omega$ established, we now estimate $f_e$-values along the streamwise extent of a gradually varied and statistically steady aerated flow down a stepped chute with $q = 0.11$ m$^2$/s, $\theta = 18.4^\circ$, step height $s = 0.06$ m, and $S_0 = \sin(\theta) = 0.32$, recorded by \cite{Bung2009}.

The GVF friction factor formulation (\ref{Eq36}) requires, in addition to these parameters, knowledge of the Favre-averaged mixture velocity $\langle \tilde{u_m} \rangle$, the Froude number $Fr$, and the equivalent clear-water depth $d_\mathrm{eq}$, along with its approximated streamwise gradient $\Delta d_\mathrm{eq}/\Delta x$. For a given flow rate $q$ and equivalent clear-water depth $d_\mathrm{eq}$, density-weighting and water-phase continuity imply $
\langle \tilde{u_m} \rangle \approx \langle \overline{u}_w \rangle = q/d_\mathrm{eq}$. This, in turn, allows the Froude number to be estimated as 

\begin{equation}
Fr =  \frac{\langle \tilde{u_m} \rangle}{\sqrt{g d_\mathrm{eq}}} \approx \frac{q}{\sqrt{g d_\mathrm{eq}^3}}. 
\end{equation}

\begin{figure}[h!]
\centering
\includegraphics{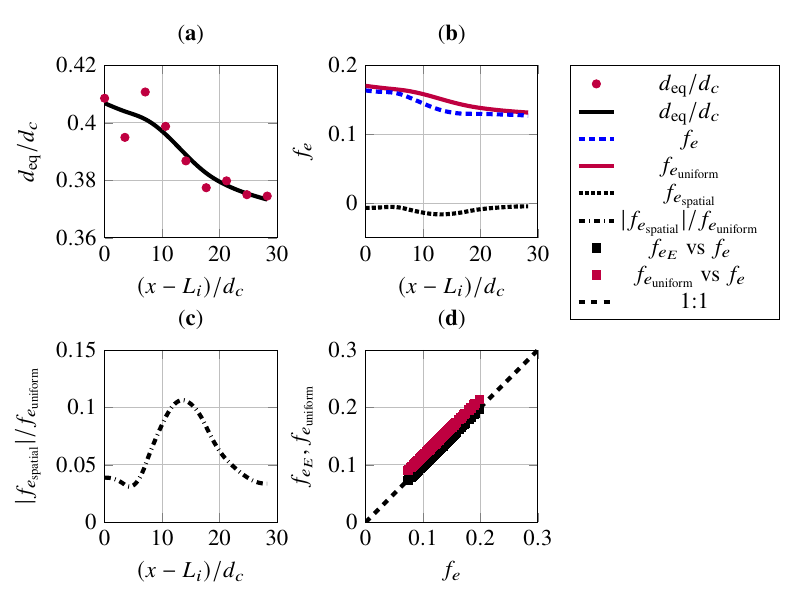}
\caption{Estimation of effective friction factors in aerated flows: (\textbf{a}) Streamwise development of $d_\mathrm{eq}/d_c$ for $q = 0.11$~m$^2$/s, $\theta = 18.4^\circ$, $s = 0.06$~m (data from \cite{Bung2009}); (\textbf{b}) Streamwise decomposition of the total friction factor $f_e$ for the same dataset; (\textbf{c}) Streamwise variation of the ratio between spatial and uniform-flow friction contributions, $|f_{e,\mathrm{spatial}}|/f_{e,\mathrm{uniform}}$, for the same dataset; (\textbf{d}) Comparison of energy-based, momentum-based, and uniform friction factor estimates, with $f_{e,E}$ and $f_{e,\mathrm{uniform}}$ plotted against $f_e$. The analysis assumes an approximated gradient of $\Delta d_\mathrm{eq}/\Delta x \approx - 2 \times 10^{-3}$.}
\label{Fig3}
\end{figure}

Figure \ref{Fig3}\textbf{a} shows the measured streamwise development of the clear-water depth along the aerated region as function of the dimensionless streamwise coordinate $(x-L_i)/d_c$, where $L_i$ denotes the upstream distance to the inception point (figure \ref{Fig0}), and $d_c = \sqrt[3]{q^2/g}$ is the critical depth. The data indicate that $d_\mathrm{eq}$ gradually decreases along the aerated region. Because the measured clear-water depths contain some noise, a smoothing spline was applied to obtain a robust estimate of the depth gradient, shown as the solid black line in figure~\ref{Fig3}\textbf{a}. Note that the maximum observed gradient was $(\Delta d_\mathrm{eq}/\Delta x)_\mathrm{max} \approx - 2 \times 10^{-3}$. 

Using the approximated gradient $\Delta d_\mathrm{eq}/\Delta x$, we now evaluate the GVF form of our explicit friction factor expression. Here, we distinguish between the total friction $f_e$, the uniform flow contribution $f_{e_\mathrm{uniform}}$, and the spatial contribution $f_{e_\mathrm{spatial}}$
\begin{equation}
 f_e =  \underbrace{\frac{ 8 \, S_0}{Fr^2} \vphantom{\left(  \beta  - \frac{\Omega \cos \theta}{Fr^2} \right)}}_{f_{e_\mathrm{uniform}}} + \underbrace{8 \, \frac{\Delta d_\mathrm{eq}}{\Delta x} \left(  \beta  - \frac{\Omega \cos \theta}{Fr^2} \right)}_{f_{e_\mathrm{spatial}}}.
\end{equation}

This decomposition allows us to isolate the contributions from uniform and spatially varying components to the total friction factor, as illustrated by the streamwise development in Figure \ref{Fig3}\textbf{b}. The spatial term is negative throughout the aerated region because flow depth decreases along the channel ($\Delta d_\mathrm{eq}/\Delta x < 0$) as the mixture accelerates. This acceleration reduces the effective friction relative to uniform flow, so that $f_e < f_{e_\mathrm{uniform}}$. To quantify this effect, figure \ref{Fig3}\textbf{c} shows the ratio $\vert f_{e_\mathrm{spatial}} \vert/ f_{e_\mathrm{uniform}}$, which reaches a maximum of approximately 0.11, indicating that spatial variation reduces the uniform-flow friction by up to 11 \%.

To evaluate differences between the momentum-based and the energy-based friction factor formulation, we calculate $f_e$ [equation (\ref{Eq36})] and $f_{e_E}$ [equation (\ref{EqFeEnergy})] for the  full dataset of \cite{Bung2009}. The measured depth profiles were noisy, so pointwise derivatives were sometimes unreliable; we therefore assume a representative gradient of $\Delta d_\mathrm{eq}/\Delta x \approx - 2 \times 10^{-3}$, in accordance with our previous findings.  This approximation is sufficient for the purpose of comparing the two formulations. Figure \ref{Fig3}\textbf{d} shows the energy-based friction factor $f_{e_E}$ plotted against the momentum based friction factor $f_e$, revealing very good agreement, with a root-mean-square  percentage deviation of 0.88 \%. For completeness, the uniform friction factor $f_{e_\mathrm{uniform}}$ is also plotted against $f_e$, yielding a substantially larger deviation of 14.0 \%. Overall, this analysis highlights the importance of accounting for spatial variations when estimating friction factors, while showing that the differences between momentum-based and energy-based formulations are comparatively small.

\section{Discussion}
\label{Sec4}

\subsection{Limitations of the proposed framework}
The present work introduces a Favre-\textcolor{black}{averaging} Shallow Water Equation (SWE) framework specifically tailored for aerated, high-Froude-number flows. By explicitly accounting for the Favre- and depth-averaged mixture continuity and momentum, the framework incorporates the effects of air entrainment, \textcolor{black}{vertically} non-uniform velocity distributions, and pressure variations within the aerated layer. Building on this SWE foundation, we derive an explicit Darcy–Weisbach friction factor formulation, in which $f_e$ is decomposed into contributions \textcolor{black}{associated with uniform steady conditions, spatial flow development, and temporal flow unsteadiness}. In addition, an equivalent energy-based GVF formulation is presented in Appendix \ref{feEnergy}, with root-mean-square percentage deviations between the two formulations found to be less than 1\% ($\S$ \ref{Sec32}).

The proposed framework provides a physically consistent means of quantifying friction in aerated flows. Despite these advantages, several limitations and assumptions should be considered:
\begin{itemize}
\item \textbf{Two-dimensionality assumption:} \textcolor{black}{The framework assumes predominantly two-dimensional flow, neglecting lateral variations in velocity and air concentration. Extension to three dimensions would require spanwise averaging of the Favre-averaged equations, introducing additional lateral stress terms.}

\item \textbf{Slowly varying correction factors:} The derivation of (\ref{Eq36}) assumes that \(\beta\) and \(\Omega\) vary slowly along the streamwise direction. In flows with abrupt changes in air concentration, or sharp accelerations, this assumption may not hold, and the full friction factor expression (\ref{eq:fe_decomp}) should be applied instead.

\item \textbf{Hydrostatic pressure assumption:} The pressure integration relies on a quasi-hydrostatic approximation, as expressed in (\ref{PressureHydro}). This assumption may become less accurate in regions with strongly curved free surfaces, intense turbulence, or significant spray, where dynamic pressure effects are non-negligible.

\item \textbf{Channel geometry effects:} The friction term is derived using the equivalent clear-water depth $d_\mathrm{eq}$. For non-rectangular channels or when sidewall corrections are needed, $d_\mathrm{eq}$ can be replaced by the hydraulic radius $R_h = A/P$, with the Froude number defined as $Fr = \langle \tilde{u_m} \rangle / \sqrt{g R_h}$, and the spatial gradient term $\Delta d_\mathrm{eq}/\Delta x$ replaced by $\Delta R_h / \Delta x$.  Here, $A$ is the cross-sectional flow area and $P$ is the wetted perimeter. This substitution allows the framework to be applied to trapezoidal or natural channels, although the relationship between spatial derivatives of $d_\mathrm{eq}$ and $R_h$ may require further calibration for complex geometries.
\end{itemize}

\subsection{Broader impact and applications}

The proposed friction factor framework carries several important implications for both fundamental research and engineering practice. First, by providing a systematic and accurate representation of friction in aerated flows, it enables improved predictions of energy dissipation in high-Froude-number spillways, chutes, and open-channel transport systems, which is essential for design, operational planning, and safety assessments.

Second, the explicit formulation is readily applicable to depth-averaged numerical solvers, such as 1D SWE models or CFD approaches, allowing aeration effects to be incorporated through the momentum and pressure correction factors 
$\beta$ and $\Omega$ without relying on ad hoc tuning parameters. This facilitates more physically consistent simulations of aerated flows across a range of conditions.

Third, the decomposition of the friction factor \textcolor{black}{into contributions associated with uniform steady conditions, spatial flow development, and temporal flow unsteadiness} offers enhanced insight into the relative importance of momentum acceleration, hydrostatic pressure gradients, and transient effects in aerated flows. Such mechanistic understanding can help identify the dominant processes in different flow regimes and improve predictive capabilities.

Finally, the framework provides a structured approach for interpreting experimental datasets, identifying conditions where classical uniform-flow formulas fail, and refining empirical or semi-empirical correction factors for both micro-rough and macro-rough inverts. This opens avenues for systematic calibration and validation of friction models in aerated flows, bridging the gap between fluid mechanics theory, laboratory experiments, and practical engineering applications.

\section{Conclusion}
\label{Sec5}
This study establishes a Favre-averaging Shallow Water Equation framework as a physically consistent basis for predicting flow resistance in aerated, high-Froude-number open-channel flows, a problem that has remained incompletely resolved in the literature. The key scientific contribution is a novel Darcy–Weisbach friction factor formulation that explicitly decomposes $f_e$ into contributions associated with uniform steady conditions, spatial flow development, and temporal flow unsteadiness, incorporating momentum and pressure correction factors ($\beta$ and $\Omega$) to account for vertically non-uniform velocity profiles and aeration-induced modifications of hydrostatic pressure.

For gradually varied, steadily developing flows, the framework reduces to a practical expression that can be directly evaluated from measured flow profiles. Application to the experimental dataset of \cite{Bung2009} demonstrates that spatial flow development systematically reduces the effective friction factor relative to the uniform-flow estimate by up to 11 \%. This finding has direct implications for the design and safety assessment of smooth and stepped spillways, where developing aerated flow regions are always present. An energy-based formulation incorporating kinetic energy and pressure correction factors ($\alpha$ and $\Omega_E$) yields nearly identical results, with deviations below 1 \%, confirming the robustness of the approach. The formulation can be extended to non-rectangular channels through the hydraulic radius, broadening its practical applicability.

From a modelling perspective, the framework is compatible with 1D SWE solvers, providing a physically based correction for aeration effects without relying on ad hoc tuning parameters. Future work should focus on detailed pressure measurements and high-resolution depth and velocity profiles to refine the correction factors and extend the framework to more transient and strongly non-uniform flows in time and space.

\medskip
\noindent\textbf{Data Availability Statement.}
Data, models, and code supporting this study are available from the corresponding author upon reasonable request.

\medskip
\noindent\textbf{Acknowledgements.}
Discussions with Dr Hanwen Cui (TMR) and Assoc Prof Stefan Felder (USNW) are acknowledged. Professor Daniel Bung (FH Aachen) is thanked for providing his dataset \textcolor{black}{and figures}. The author acknowledges the use of the AI language model ChatGPT (OpenAI) for editorial assistance. All derivations, interpretations, and conclusions in this work are solely those of the author.

\medskip
\noindent\textbf{Declaration of Interest.}
The author declares no conflict of interest.

\section*{Nomenclature}

\subsection*{Latin symbols}
\begin{description}[leftmargin=2.8cm, labelwidth=1cm, align=left]
  \item[$A$] Cross-sectional flow area (m$^2$)
  \item[$c$] Volumetric air concentration (--)
  \item[$d$] Single-phase (water) flow depth (m)
  \item[$d_\mathrm{eq}$] Equivalent clear-water flow depth (m)
  \item[$e_m$] Mechanical energy of the mixture per unit volume (Pa = J/m$^3$)
  \item[$F_e$] Local energy flux of the mixture per unit area (W/m$^2$)
  \item[$f_e$] Equivalent friction factor (--)
  \item[$Fr$] $= \langle \tilde{u}_m \rangle / \sqrt{g d_\mathrm{eq}}$; Froude number (--)
  \item[$g$] Gravitational acceleration (m/s$^2$)
  \item[$g_x$] $= g S_0$; streamwise component of gravitational acceleration (m/s$^2$)
  \item[$H_m$] Mean total head of the mixture, referenced to the channel bed $z_0$ (m)
  \item[$L_i$] Upstream distance to aeration inception (m)
  \item[$P$] Wetted perimeter (m)
  \item[$p$] Pressure (N/m$^2$)
  \item[$q$] Specific water discharge (m$^2$/s)
  \item[$R_h$] $= A/P$; hydraulic radius (m)
  \item[$s$] Step height (m)
  \item[$S_0$] $= -\partial z_0/\partial x$ or $\sin\theta$; bed slope (--)
  \item[$S_f$] Friction slope (--)
  \item[$t$] Time (s)
  \item[$u_m$] Streamwise mixture velocity (m/s)
  \item[$u_w$] Streamwise water velocity (m/s)
  \item[$w_m$] Vertical mixture velocity (m/s)
  \item[$x$] Streamwise coordinate measured along the channel bed (m)
  \item[$z_0$] Channel bed elevation, measured vertically from a fixed datum (m)
    \item[$z$] Bed-normal coordinate, measured positive upward from the channel bed (m)
  \item[$z_{90}$] Mixture depth at which $\overline{c} = 0.9$ (m)
  \item[$\langle \tilde{z} \rangle$] Favre centroid of the mixture (m)
\end{description}

\subsection*{Greek symbols}
\begin{description}[leftmargin=2.8cm, labelwidth=1cm, align=left]
  \item[$\alpha$] Kinetic energy correction factor (--)
  \item[$\beta$] Momentum correction factor (--)
  \item[$\Omega$] Momentum-based pressure correction factor (--)
  \item[$\Omega_E$] Energy-based pressure correction factor (--)
  \item[$\rho$] Density (kg/m$^3$)
  \item[$\tau_0$] Bed shear stress (N/m$^2$)
  \item[$\theta$] Bed slope angle (rad)
  \item[$\phi$] Arbitrary quantity (e.g.\ $u_m$)
  \item[$\phi''$] Favre fluctuation of $\phi$ (same units as $\phi$)
\end{description}

\subsection*{Indices and operators}
\begin{description}[leftmargin=2.8cm, labelwidth=1cm, align=left]
  \item[$a$] Air phase
  \item[$aw$] Air--water interface
  \item[\normalfont  GVF] Gradually Varied Flow (steady, spatially developing flow)
  \item[\normalfont  max] Maximum
  \item[$m$] Mixture
  \item[\normalfont  SWE] Shallow Water Equations
  \item[$w$] Water phase
  \item[$\overline{(\cdot)}$] Time-averaging operator; e.g.\ $\overline{u}$
  \item[$\tilde{(\cdot)}$] Favre (density-weighted) averaging operator;
  $\tilde{u}_m = \overline{\rho_m u_m}/\overline{\rho_m}$
  \item[$\langle \cdot \rangle$] Depth-averaging operator;
  $\langle \overline{c} \rangle = \frac{1}{z_{90}} \int_{z=0}^{z_{90}} \overline{c}\,\mathrm{d}z$
  \item[$\langle \tilde{(\cdot)} \rangle$] Favre- and depth-averaging operator;
  $\langle \tilde{u}_m \rangle =
  \frac{\int_{z=0}^{z_{90}} \overline{\rho_m u_m}\,\mathcal{d}z}
       {\int_{z=0}^{z_{90}} \overline{\rho_m}\,dz}$
\end{description}

\begin{appendix}
\section{Momentum-based friction factor formulation}
\label{AppendixA}
The density-weighted momentum equation of the Shallow Water Equations (SWE) for aerated flows follows from (\ref{Eq2.15}) and is written as

\begin{equation}
\frac{\partial (d_\mathrm{eq} \langle \tilde{u_m} \rangle)}{\partial t} 
+ \frac{\partial (\beta \, d_\mathrm{eq} \langle \tilde{u_m} \rangle^2)}{\partial x} 
= g d_\mathrm{eq} S_0 - \frac{\partial}{\partial x} \left( \frac12 g \, \Omega \, d_\mathrm{eq}^2 \cos\theta \right) - \frac{\tau_0}{\rho_w},
\label{A1}
\end{equation}
where $d_\mathrm{eq}$ is the equivalent clear-water depth, $\langle \tilde{u_m} \rangle$ is the Favre- and depth-averaged water velocity, $\tau_0$ is the bed shear stress, $\beta$ is the momentum correction factor, and $\Omega$ is the pressure correction factor. The bed shear stress is expressed in Darcy-Weisbach form as

\begin{equation}
\tau_0 = \frac{1}{8} \, \rho_w \, f_e \, \langle \tilde{u_m} \rangle^2,
\end{equation}
which can be substituted into (\ref{A1}), yielding

\begin{equation}
\frac{\partial (d_\mathrm{eq} \langle \tilde{u_m} \rangle)}{\partial t} 
+ \frac{\partial (\beta \, d_\mathrm{eq} \langle \tilde{u_m} \rangle^2)}{\partial x} 
= g d_\mathrm{eq} S_0 - \frac{\partial}{\partial x} \left( \frac12 g \, \Omega \, d_\mathrm{eq}^2 \cos\theta \right) - \frac{1}{8} f_e \langle \tilde{u_m} \rangle^2.
\end{equation}

Rearranging for $f_e$ and replacing the derivatives with finite differences gives
\begin{equation}
f_e = \frac{8}{\langle \tilde{u_m} \rangle^2} 
\left(
g\,d_\mathrm{eq} S_0
- \frac{\Delta \left( \frac12 g\,\Omega\,d_\mathrm{eq}^2 \cos\theta \right)}{\Delta x}
- \frac{\Delta (\beta\,d_\mathrm{eq}\langle \tilde{u_m} \rangle^2)}{\Delta x}
- \frac{\Delta (d_\mathrm{eq} \langle \tilde{u_m} \rangle)}{\Delta t}
\right),
\label{feFull}
\end{equation}
which is the most general conservative form of the friction factor expression that directly follows from the SWE momentum equation. It applies to arbitrary unsteady and spatially non-uniform aerated flows, including cases with strong variations in velocity, depth, or aeration. All the parameters in the equation can in principle be measured or estimated; however, evaluating the derivatives in particular can be cumbersome in practice, especially in highly non-uniform or rapidly varying flows.

To simplify (\ref{feFull}) and express it in a Gradually Varied Flow (GVF) framework, we assume steady flow, i.e., $\frac{\Delta (d_\mathrm{eq} \langle \tilde{u_m} \rangle)}{\Delta t} \approx 0$. Further, we postulate the $\beta$ and $\Omega$ vary slowly with $x$, which leads to the following simplifications of the spatial derivatives

\begin{align}
\frac{\Delta \left( \frac12 g \, \Omega \, d_\mathrm{eq}^2 \cos\theta \right)}{\Delta x}  &\approx g d_\mathrm{eq} \frac{\Delta d_\mathrm{eq}}{\Delta x} \, \Omega \cos\theta,\\
\frac{\Delta (\beta \, d_\mathrm{eq} \langle \tilde{u_m} \rangle^2)}{\Delta x} &\approx \beta \frac{\Delta (d_\mathrm{eq} \langle \tilde{u_m} \rangle^2)}{\Delta x} = -\beta g d_\text{eq} Fr^2 \frac{\Delta d_\text{eq}}{\Delta x},
\label{convective}
\end{align}
where  the Froude number is defined as $
Fr = \langle \tilde{u_m} \rangle /\sqrt{g \, d_\mathrm{eq}}$. The minus sign in (\ref{convective}) arises from applying the steady 1D continuity equation, \(
d_\mathrm{eq} \, \langle \tilde{u_m} \rangle = q, \) 
which allows us to express the derivative of \(d_\mathrm{eq} \, \langle \tilde{u_m} \rangle^2\) in terms of the derivative of \(d_\mathrm{eq}\). Substituting these expressions into (\ref{feFull}) yields a novel expression for the GVF friction factor

\begin{equation}
f_e=\frac{8}{Fr^2}
\left(
S_0
+ \frac{\Delta d_\mathrm{eq}}{\Delta x}
\left(\beta\,Fr^2-\Omega\cos\theta\right)
\right) =  \underbrace{\frac{ 8 \, S_0}{Fr^2} \vphantom{\left(  \beta  - \frac{\Omega \cos \theta}{Fr^2} \right)}}_{f_{e_\mathrm{uniform}}} + \underbrace{8 \, \frac{\Delta d_\mathrm{eq}}{\Delta x} \left(  \beta  - \frac{\Omega \cos \theta}{Fr^2} \right)}_{f_{e_\mathrm{spatial}}}.
\label{feMomentum}
\end{equation}

This formulation captures the contributions of convective momentum flux and pressure forces in a steady, gradually varied flow. By assuming slowly varying correction factors, it provides a practical expression for friction in aerated channels while still retaining the effects of \textcolor{black}{vertically} non-uniform velocity profiles (\(\beta\)) and aeration-induced pressure correction (\(\Omega\)). The overall structure of this expression is similar in form to single-phase formulations (e.g., \cite{Graf1995}).

\section{Energy-based friction factor formulation}
\label{feEnergy}
We consider an aerated, high-Froude-number flow in a sloping channel with bed slope $S_0 = -\partial z_0 / \partial x$. Let $\tilde{u_m}(z)$ denote the depth-dependent Favre-averaged mixture velocity, where $x$ is aligned with the channel and $z$ is the bed-normal direction (figure \ref{Fig0}). Using Favre averaging, the local energy flux of the mixture is written as

\begin{equation}
F_e(z) = \tilde{u_m} \, e_m(z),
\end{equation}
where $F_e$ is the energy flux per unit area and $e_m(z)$ is the mechanical energy of the mixture per unit volume. The mechanical energy consists of kinetic, potential, and pressure contributions and is given by

\begin{equation}
e_m(z) = \frac{1}{2} \overline{\rho_m} \tilde{u_m}^2 + \overline{\rho_m} g z \cos\theta + \overline{p_m},
\end{equation}
where $\overline{\rho_m}$ and $\overline{p_m}$ are the time-averaged mixture density and pressure, respectively. The former is approximated as 
$\overline{\rho_m} \approx \rho_w (1-\overline{c})$, where 
$\overline{c}$ is the local air concentration and $\rho_w$ is the water density. This approximation neglects air density, which is appropriate for high-speed aerated flows. Integrating the local energy flux over the flow depth yields the total energy flux per unit width, $\int_{z=0}^{z_{90}} F_e(z) \, \mathrm{d}z$. This integral can be decomposed into kinetic and gravitational (potential plus pressure) components, such that

\begin{equation}
\int_{z=0}^{z_{90}} F_e(z) \, \mathrm{d}z
= \int_{z=0}^{z_{90}} \frac{1}{2} \overline{\rho_m} \tilde{u_m}^3 \mathrm{d}z
+ \int_{z=0}^{z_{90}} \tilde{u_m} \, \overline{\rho_m} g z \cos\theta \,\mathrm{d}z + \int_{z=0}^{z_{90}} \tilde{u_m} \, \overline{p_m} \, \mathrm{d}z.
\end{equation}

The depth-integrated kinetic energy flux can be approximated using the equivalent clear-water depth $d_\mathrm{eq}$ and the depth-averaged Favre velocity $\langle \tilde{u_m} \rangle$ as

\begin{equation}
\int_{z=0}^{z_{90}} \frac{1}{2} \overline{\rho_m} \tilde{u_m}^3 \mathrm{d}z
= \frac{1}{2} \rho_w \int_{z=0}^{z_{90}} (1-\overline{c}) \tilde{u_m}^3 \mathrm{d}z
= \frac{1}{2} \rho_w \alpha \, d_\mathrm{eq} \, \langle \tilde{u_m} \rangle^3,
\end{equation}
where the kinetic energy correction factor

\begin{equation}
\alpha = \frac{\langle \tilde{u_m}^3 \rangle}{\langle \tilde{u_m} \rangle^3}
= \frac{\int_{z=0}^{z_{90}} (1-\overline{c}) \tilde{u_m}^3 \mathrm{d}z}{d_\mathrm{eq} \langle \tilde{u_m} \rangle^3}
\label{EqAlpha}
\end{equation}
accounts for the vertical non-uniformity of the mixture velocity and aeration distribution. The gravitational energy flux includes both potential and pressure contributions. Assuming a hydrostatic pressure distribution based on the local mixture density, the depth-integrated gravitational energy flux can be written as

\begin{equation}
\int_{z=0}^{z_{90}} \tilde{u_m} (\overline{\rho_m} g z \cos\theta + \overline{p_m}) \, \mathrm{d}z
= \rho_w g  \cos\theta \, \Omega_E \, d_\mathrm{eq}^2 \, \langle \tilde{u_m} \rangle,
\end{equation}
where the energy-based pressure correction factor

\begin{equation}
\Omega_E = \frac{\int_{z=0}^{z_{90}} \tilde{u}_m \left((1 - \overline{c}) \, z + \int_z^{z_{90}} (1 - \overline{c}(\zeta)) \, \mathrm{d}\zeta \right) \mathrm{d}z}{d_\mathrm{eq}^2 \, \langle \tilde{u}_m \rangle} 
\label{EqOmegaE}
\end{equation}
captures the effects of the aeration-induced pressure distribution and the velocity profile. Combining kinetic and gravitational contributions, the total depth-integrated energy flux reads

\begin{equation}
\int_{z=0}^{z_{90}} F_e(z) \, \mathrm{d}z
= \frac{1}{2} \rho_w \alpha \, d_\mathrm{eq} \, \langle \tilde{u_m} \rangle^3
+ \rho_w g \cos\theta \, \Omega_E \, d_\mathrm{eq}^2 \, \langle \tilde{u_m} \rangle ,
\end{equation}
which, when normalized by $\rho_w g d_\mathrm{eq} \langle \tilde{u_m} \rangle$, defines the flux-averaged total head of the mixture

\begin{equation}
H_m = \frac{1}{\rho_w g d_\mathrm{eq} \langle \tilde{u_m} \rangle} \int_{z=0}^{z_{90}} F_e(z) \, \mathrm{d}z
= \alpha \frac{\langle \tilde{u_m} \rangle^2}{2 g} + \Omega_E d_\mathrm{eq} \cos\theta.
\end{equation}

Here, the flux-averaged total head $H_m$ is referenced to the channel bed elevation $z_0$ (figure \ref{Fig0}), to maintain consistency with the momentum-based derivation. The friction slope follows from (\ref{EqSF})

\begin{equation}
S_f = S_0 - \frac{\partial H_m}{\partial x} 
= S_0 - \frac{\partial}{\partial x} \Bigg( \alpha \frac{\langle \tilde{u_m} \rangle^2}{2 g} + \Omega_E d_\mathrm{eq} \cos \theta \Bigg),
\end{equation}
and the corresponding Darcy–Weisbach friction factor is

\begin{equation}
f_{e_E} = \frac{8 g d_\mathrm{eq}}{\langle \tilde{u_m} \rangle^2} S_f
= \frac{8 g d_\mathrm{eq}}{\langle \tilde{u_m} \rangle^2} 
\left( S_0 - \frac{\partial}{\partial x} \Big( \alpha \frac{\langle \tilde{u_m} \rangle^2}{2 g} + \Omega_E d_\mathrm{eq} \cos \theta \Big) \right).
\label{B11}
\end{equation}

For gradually varied flows, where streamwise gradients of $\alpha$ and $\Omega_E$ are small and 1D continuity holds ($d_\mathrm{eq} \langle \tilde{u_m} \rangle = q$), (\ref{B11}) can be simplified

\begin{equation}
f_{e_E} = \frac{8}{Fr^2} \left( S_0 + \frac{\Delta d_\mathrm{eq}}{\Delta x} \left( \alpha \, Fr^2 - \Omega_E \cos\theta \right) \right) = \underbrace{\frac{ 8 \, S_0}{Fr^2} \vphantom{\left(  \beta  - \frac{\Omega \cos \theta}{Fr^2} \right)}}_{f_{e_\mathrm{uniform}}} + \underbrace{8 \, \frac{\Delta d_\mathrm{eq}}{\Delta x} \left(  \alpha  - \frac{\Omega_E \cos \theta}{Fr^2} \right)}_{f_{e_\mathrm{spatial}}},
\label{EqFeEnergy}
\end{equation}
where the derivatives are expressed in finite-difference form, and the local Froude number is $Fr = \langle \tilde{u_m} \rangle / \sqrt{g d_\mathrm{eq}}$. 

The energy-based friction factor is closely aligned with the corresponding momentum-based formulation. Differences arise naturally from the flux-weighted Favre averaging inherent to the energy balance, since the energy flux contains one additional factor of velocity relative to the momentum flux. As a result, regions of higher velocity contribute disproportionately to the energy flux. These effects are accounted for by the kinetic energy correction factor
 $\alpha$ and the energy-based pressure correction factor $\Omega_E$. For nearly uniform velocity profiles, $\tilde{u_m} \approx \langle \tilde{u_m} \rangle$, such that $\alpha \approx \beta \approx 1$ and $\Omega_E \approx \Omega$. In this limiting case, the momentum-based and energy-based friction formulations become equivalent.

\end{appendix}

\bibliographystyle{jfm}
\bibliography{bibliography}

\end{document}